\documentclass[letterpaper, 10 pt, conference]{ieeeconf}
\IEEEoverridecommandlockouts
\usepackage[USenglish]{babel}
\usepackage{xcolor,graphics,graphicx}
\usepackage{epsfig} 
\usepackage{amsmath} 
\usepackage{amssymb}  
\usepackage{amsfonts}
\usepackage{mathtools}
\usepackage{optidef,booktabs}
\usepackage{cite}
\usepackage{url}
\usepackage{lipsum}
\usepackage{hyperref}
\hypersetup{hidelinks}
\usepackage[utf8]{inputenc}
\usepackage[font=small,skip=0pt]{subcaption}
\usepackage[font=small,skip=0pt]{caption}
\usepackage{tabularray}
\usepackage{tikz}
\usetikzlibrary{decorations.markings}

\UseTblrLibrary{amsmath}

\newtheorem{remark}{Remark}

\newtheorem{lemma}{Lemma}
\newtheorem{theorem}{Theorem}
\newtheorem{corollary}{Corollary}

\title{\LARGE \bf Feedback Stability of Discrete-Time LTI Systems via Frequency-Wise Scaled Relative Graphs}

\author{Guoming Shi$^{1}$, Eder Baron-Prada$^{2}$ and Fredy Ruiz$^{1}$
\thanks{*This work was partially supported by the China Scholarship Council.}%
\thanks{$^{1}$Guoming Shi and Fredy Ruiz are with the Dipartimento di Elettronica, Informazione e Bioingegneria, Politecnico di Milano, 20133 Milan, Italy {\tt\small \{guoming.shi, fredy.ruiz\}@polimi.it}}%
\thanks{$^{2}$Eder Baron-Prada is with the Austrian Institute of Technology, 1210 Vienna, Austria {\tt\small eder.baron-prada@ait.ac.at}}%
}

\begin{document}

\maketitle
\thispagestyle{empty}
\pagestyle{empty}

\begin{abstract}
This paper studies the stability of negative-feedback interconnections of proper,
stable discrete-time multivariable linear time-invariant (LTI) systems using scaled relative graphs (SRGs).
We propose a frequency-wise condition that compares the SRGs of the two subsystems'
frequency-response matrices only at the same frequency along a feedback-scaling homotopy.
We prove that this condition guarantees well-posedness along the homotopy and $\ell_2$-stability
for both single-input single-output (SISO) and multi-input multi-output (MIMO) subsystems,
without requiring the chord property or nonsingular frequency-response matrices.
An equivalent form exchanges scaling and geometric inversion.
We also derive an operator-level condition from the SRG closures of the convolution operators
and show that it implies the frequency-wise condition.
Numerical examples show that the operator-level condition is more conservative 
because it depends on intersections between responses at different frequencies.
\end{abstract}

\section{Introduction}
SRGs were introduced by Ryu et al.~\cite{ryu2022scaled} to analyze the convergence of optimization algorithms.
Chaffey et al.~\cite{chaffey2023} then applied SRGs to nonlinear system analysis, 
where feedback stability can be checked by verifying the separation of the subsystems' SRGs.

For LTI systems, frequency-domain analysis is an important part of classical feedback stability analysis.
The Nyquist criterion determines closed-loop stability from the open-loop frequency response~\cite{zhou1998essentials}.
The small-gain theorem~\cite{zames1966input} requires only a gain bound for each subsystem 
and guarantees stability against norm-bounded uncertainty.
For MIMO systems,
the small-phase theorem adds phase information to complement the small-gain theorem~\cite{chen2024phase}.

These theorems treat gain and phase separately; however, SRGs can handle both in a single graph. 
Building on the SRG separation theorem of~\cite{chaffey2023}, recent work has advanced SRG-based stability analysis in both theory and applications. 
Theoretically, a homotopy argument corrects two assumptions of the original separation theorem~\cite{chaffey2025homotopy}, 
and soft and hard SRGs remove the chord assumption and admit possibly unbounded open-loop systems~\cite{chen2026soft}. 
Multiplier-defined containment regions further allow certifying hard stability from soft containment, 
without a homotopy condition~\cite{baron2026Equivalence}. 

For MIMO LTI systems, 
frequency-wise SRG conditions certify closed-loop stability while analyzing each subsystem separately~\cite{baron2025stability}, 
and SRGs yield a small-phase theorem without the sectoriality assumption~\cite{baron2025mixed}. 
For nonlinear systems, 
stability results have been derived for Lur'e systems with unstable linear plants through a generalized circle criterion~\cite{krebbekx2025lure}, 
for general interconnections of SISO nonlinear systems~\cite{krebbekx2025general}, 
and for non-square MIMO  systems~\cite{krebbekx2025nonlinear}. 
For nonlinear MIMO systems, weighted scaled graphs exploit system structure to reduce conservatism~\cite{degroot2025exploiting}. 
In applications, scaled graphs have been used to analyze and design reset control systems~\cite{degroot2026reset}, 
derive decentralized stability conditions for multi-agent systems~\cite{baron2025decentralized}, 
and assess the stability of power-electronics-dominated grids~\cite{baron2026grids}.

In data-driven control, controllers are designed directly from data rather than from a model,
and certifying closed-loop stability without an accurate model remains difficult. 
SRGs are promising for this purpose 
because the SRG of a discrete-time LTI system can be computed directly from input--output data, 
and a robust SRG computed from noisy data is guaranteed to contain the true system's SRG~\cite {nauta2026Discrete-time}. 
However, existing work on discrete-time SRGs has focused on their computation~\cite{nauta2026Discrete-time,shi2026computing}, 
and SRG-based stability conditions for discrete-time feedback interconnections have received little attention.

To fill this gap and provide a basis for stability analysis of data-driven controllers, we first revisit the continuous-time frequency-wise condition of~\cite{baron2025stability}
and derive an equivalent form that exchanges scaling and geometric inversion between the two subsystem SRGs;
the inversion acts on SRG sets and requires no inverse subsystem realization.
We then establish a frequency-wise SRG condition for proper, stable discrete-time LTI systems,
covering both SISO and MIMO cases.
The condition compares the subsystem SRGs only at the same frequency
and guarantees well-posedness along a feedback-scaling homotopy and $\ell_2$-stability,
without requiring the chord property or nonsingular frequency-response matrices.
Finally, we derive an operator-level condition from the SRG closures of the convolution operators
and show that it implies the frequency-wise condition.
Hence, the frequency-wise condition is less conservative than the operator-level condition,
which can fail because of intersections between responses at different frequencies, as illustrated by SISO and MIMO examples.

The paper is organized as follows.
Section~\ref{Preliminaries} introduces the notation and preliminaries on SRGs.
Section~\ref{Stability} presents the frequency-wise SRG stability conditions.
Section~\ref{OperatorLevel} derives the operator-level condition.
Section~\ref{Examples} gives numerical examples, and Section~\ref{Conclusions} presents the conclusions.

\section{Preliminaries}\label{Preliminaries}
\subsection{Notation}
Let $\mathbb{N}_0:=\{0,1,2,\ldots\}$, and let $j$ denote the imaginary unit.
For $z\in\mathbb{C}$, $\bar z$, $\Re(z)$, $\Im(z)$, and $|z|$ denote its complex conjugate,
real part, imaginary part, and modulus, respectively.
For a matrix $A$, $A^*$ denotes its conjugate transpose,
and $\det(A)$ denotes its determinant when $A$ is square.
The symbol $I$ denotes the identity matrix or operator, as appropriate.
$\mathcal{RH}_\infty^{m\times m}$ denotes the set of $m\times m$ proper, real-rational,
and stable transfer matrices,
where stability means that all poles lie in the open left half-plane in continuous time
and in the open unit disk in discrete time.
For $\mathcal{S}\subseteq\mathbb{C}$ and $\alpha\in\mathbb{R}$,
let $\alpha\mathcal{S}:=\{\alpha z:z\in\mathcal{S}\}$.
The closure and the convex hull of $\mathcal{S}$ are denoted by $\mathrm{cl}\,\mathcal{S}$
and $\mathrm{co}\,\mathcal{S}$, respectively.
The extended complex plane is $\widehat{\mathbb{C}}:=\mathbb{C}\cup\{\infty\}$.
For $\mathcal{S}\subseteq\widehat{\mathbb{C}}$, define its geometric inverse by
$\mathcal{S}^{-1}:=\{1/\bar z:z\in\mathcal{S}\}$, with $0\mapsto\infty$ and $\infty\mapsto0$.
Nonzero scalings leave $\infty$ fixed.

\subsection{Signals and Systems}
Let $\mathcal{L}_2:=\mathcal{L}_2([0,\infty);\mathbb{C}^m)$ denote the space of square-integrable signals.
Throughout the paper, 
$\ell_2:=\ell_2(\mathbb{N}_0;\mathbb{C}^m)$ denotes the space of one-sided square-summable sequences.
Its inner product is $\langle u,y\rangle:=\sum_{k=0}^\infty u_k^*y_k$, 
with induced norm $\|u\|:=\langle u,u\rangle^{1/2}$.
A discrete-time system $H\in\mathcal{RH}_\infty^{m\times m}$ induces a bounded causal
convolution operator $T_d:\ell_2\to\ell_2$ with zero initial conditions,
whose frequency response is $H(e^{j\theta})$, $\theta\in[0,2\pi)$.
The direct term of $H$ is $D:=\lim_{z\to\infty}H(z)$.

\subsection{Scaled Relative Graphs}
Let $\mathcal{X}$ be a Hilbert space with inner product $\langle\cdot,\cdot\rangle$.
The angle between nonzero $x,y\in\mathcal{X}$ is
$\angle(x,y):=\arccos\big(\Re\langle x,y\rangle/(\|x\|\,\|y\|)\big)\in[0,\pi]$.
The SRG of an operator $H$ on $\mathcal{X}$ is~\cite{ryu2022scaled,chaffey2023}
\begin{align*}
    \operatorname{SRG}(H):=\left\{\frac{\|y_1-y_2\|}{\|u_1-u_2\|}\,e^{\pm j\angle(u_1-u_2,\,y_1-y_2)}\right\},
\end{align*}
where $u_1\neq u_2$ range over $\mathcal{X}$ and $y_i=Hu_i$; if $y_1=y_2$,
the angle is taken to be zero.
For linear $H$, it suffices to take $u_2=0$.
For a matrix $M\in\mathbb{C}^{m\times m}$,
$\operatorname{SRG}(M)$ denotes the SRG of the linear map $x\mapsto Mx$ on $\mathbb{C}^m$.

We use the following SRG identities~\cite{chaffey2023}.
For any $\alpha\in\mathbb{R}$,
\begin{align*}
    \operatorname{SRG}(\alpha H) = \alpha\,\operatorname{SRG}(H).
\end{align*}
In particular, $\operatorname{SRG}(-H) = -\operatorname{SRG}(H)$.
If $H^{-1}$ exists, then the SRG satisfies the inverse rule
\begin{align*}
    \operatorname{SRG}(H^{-1})  = \operatorname{SRG}(H)^{-1}.
\end{align*}
A set $\mathcal{S}\subseteq\mathbb{C}$ satisfies the chord property if
\begin{align*}
    z\in\mathcal{S}
    \quad\Longrightarrow\quad
    [z,\bar z]\subseteq\mathcal{S},
\end{align*}
where $[z_1,z_2]:=\{\lambda z_1+(1-\lambda)z_2:\lambda\in[0,1]\}$.
An operator satisfies the chord property if its SRG does.

\subsection{SRGs of Matrices}
We denote the Beltrami--Klein map as follows~\cite{pates2021scaled},
\begin{align*}
    f_{bk}(z)=\frac{(\bar z-j)(z-j)}{1+|z|^2}=\frac{|z|^2-1-2j\,\Re(z)}{1+|z|^2},
\end{align*}
which maps $z$ and $\bar z$ to the same point of the closed unit disk.
Its inverse on conjugate-symmetric sets is
\begin{align*}
    g_{bk}(w)=\left\{\frac{\Im(w) \pm j\sqrt{1-|w|^2}}{\Re(w) - 1}\right\},\qquad |w|\leq 1,\ w\neq 1.
\end{align*}
For $M\in\mathbb{C}^{m\times m}$, let $P_M:=I+M^*M$, which is positive definite,
and define the matrix transform
\begin{align*}
    f_{bk}(M):=P_M^{-1/2}(M^*-jI)(M-jI)P_M^{-1/2}.
\end{align*}
The numerical range of $M$ is
$\mathcal{W}(M):=\{x^*Mx:x\in\mathbb{C}^m,\ \|x\|=1\}$~\cite{psarrakos2002numerical}.
The following lemma shows how the SRG of a frequency-response matrix is computed.

\begin{lemma}[Matrix SRGs~\cite{pates2021scaled}]\label{lem:matrix-srg}
    For every $M\in\mathbb{C}^{m\times m}$, $f_{bk}(\operatorname{SRG}(M))=\mathcal{W}(f_{bk}(M))$.
    Consequently, $\operatorname{SRG}(M)=g_{bk}(\mathcal{W}(f_{bk}(M)))$.
\end{lemma}

\section{Frequency-Wise SRG Stability Conditions}\label{Stability}
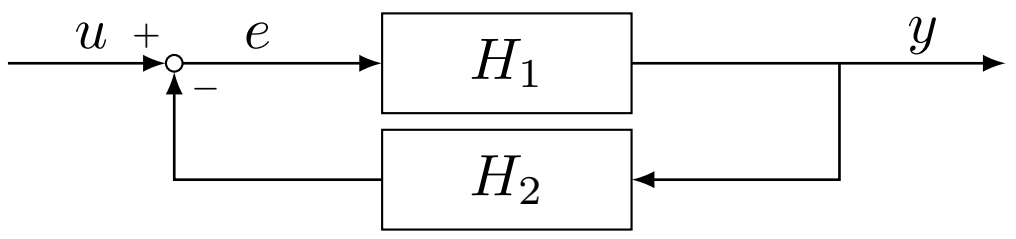
\begin{figure}[hbt]
    \centering
    \begin{tikzpicture}[scale=1.1, every node/.style={transform shape}]
    \draw (9.75,4.3) rectangle (11.25,3.7);
    \node at (10.5,4) {$H_1$};
    \draw (9.75,3.6) rectangle (11.25,3);
    \node at (10.5,3.3) {$H_2$};
    \draw[-latex, line width = .5 pt] (12.5,4) -- (13.5,4);
    \draw[-latex, line width = .5 pt] (11.25,4) -- (12.5,4) -- (12.5,3.3) -- (11.25,3.3);
    \draw[fill = white] (8.5,4) circle [radius=0.05];
    \draw[-latex, line width = .5 pt] (9.75,3.3) -- (8.5,3.3) -- (8.5,3.95) ;
    \draw[-latex, line width = .5 pt] (8.55,4) -- (9.75,4);
    \draw[-latex, line width = .5 pt] (7.5,4) -- (8.45,4);
    \node at (8.6892,3.8444) {\tiny$-$};
    \node at (8.3332,4.1664) {\tiny$+$};
    \node at (9,4.1664) {$e$};
    \node at (8,4.1664) {$u$};
    \node at (13,4.1664) {$y$};
    \end{tikzpicture}
    \vspace{5pt}
    \caption{Feedback interconnection between $H_1$ and $H_2$.}
    \label{fig:feedback_system_diagram}
\end{figure}

This section establishes frequency-wise SRG stability conditions for discrete-time LTI feedback systems.
Consider the negative-feedback interconnection in Fig.~\ref{fig:feedback_system_diagram}.
We first recall a continuous-time frequency-wise result for comparison.
We then prove the discrete-time result directly.
Its connection to operator SRG closures is discussed in Section~\ref{OperatorLevel}.

For a zero-initial-state LTI interconnection,
input--output stability means that the closed-loop map is causal and bounded on the
corresponding $\mathcal{L}_2$ or $\ell_2$ signal space~\cite{zhou1998essentials}.
For proper rational subsystems,
algebraic well-posedness at a homotopy parameter $\tau$ means $\det(I+\tau D_1D_2)\ne0$,
where $D_i$ is the direct term of $H_i$.

\begin{lemma}[Frequency-wise SRG separation~\cite{baron2025stability}]\label{GFT for LTI systems}
    Consider the feedback interconnection shown in Fig.~\ref{fig:feedback_system_diagram}.
    Assume that $H_1(s), H_2(s) \in \mathcal{RH}^{m\times m}_\infty$ are continuous-time LTI systems,
    and that either $H_1(s)$ or $H_2(s)$ satisfies the chord property.
    If, for all $\omega \in [0,\infty)$ and all $\tau\in(0,1]$,
    \begin{align*}
        \operatorname{SRG}(H_1(j\omega))^{-1} \cap-\tau\operatorname{SRG}(H_2(j\omega))=\varnothing,
    \end{align*}
    then the closed-loop system is $\mathcal{L}_2$-stable.
\end{lemma}

According to the SRG rules of operation and Lemma~\ref{GFT for LTI systems},
the following corollary holds.
\begin{corollary}\label{GFT inv for LTI systems}
    Consider the feedback interconnection shown in Fig.~\ref{fig:feedback_system_diagram}.
    Assume that $H_1(s), H_2(s) \in \mathcal{RH}^{m\times m}_\infty$,
    and that either $H_1(s)$ or $H_2(s)$ satisfies the chord property.
    If, for all $\omega \in [0,\infty)$ and all $\tau\in(0,1]$,
    \begin{align}
        \tau\operatorname{SRG}(H_1(j\omega)) \cap -\operatorname{SRG}(H_2(j\omega))^{-1} = \varnothing,
    \end{align}
    then the closed-loop system is $\mathcal{L}_2$-stable.
\end{corollary}
\begin{proof}
    Fix arbitrary $\omega\geq0$ and $\tau\in(0,1]$, and write
    $\mathcal{S}_i:=\operatorname{SRG}(H_i(j\omega))$, $i=1,2$.
    By the condition of Corollary~\ref{GFT inv for LTI systems},
    \begin{align*}
        \tau\mathcal{S}_1 \cap -\mathcal{S}_2^{-1} = \varnothing.
    \end{align*}
    Geometric inversion $z\mapsto1/\bar z$ is a bijection on $\widehat{\mathbb{C}}$ and preserves disjointness.
    Applying inversion to both sets gives
    \begin{align*}
        (\tau\mathcal{S}_1)^{-1} &\cap (-\mathcal{S}_2^{-1})^{-1} = \varnothing,\\
        \frac{1}{\tau}\mathcal{S}^{-1}_1 &\cap -\mathcal{S}_2 = \varnothing.
    \end{align*}
    Multiplying both sets by $\tau > 0$ yields
    \begin{align*}
        \mathcal{S}_1^{-1} \cap -\tau\mathcal{S}_2 &= \varnothing.
    \end{align*}
    Since $\omega$ and $\tau$ were arbitrary, Lemma~\ref{GFT for LTI systems} gives $\mathcal{L}_2$-stability.
\end{proof}

The equivalent form in Corollary~\ref{GFT inv for LTI systems} can be useful when its SRG sets are easier to construct or separate than those in Lemma~\ref{GFT for LTI systems}.
In both forms, the inverse notation denotes the geometric inversion of an SRG set and requires no inverse-system realization.
For example, the inverse of a strictly proper SISO transfer function is improper,
whereas a proper and stable SISO inverse requires a biproper, minimum-phase transfer function.
Neither separation condition imposes these requirements on an inverse transfer function.

\begin{remark}\label{rem:integrator}
    The equivalent formulation in Corollary~\ref{GFT inv for LTI systems} retains the assumption that the original continuous-time subsystems are proper and stable.
    In particular, an uncancelled integrator with a pole at $s=0$ lies outside this system class.
    Geometric inversion of an SRG does not make the criterion applicable to an unstable or unbounded subsystem.
    Such cases require separate analysis, for example, through a suitable loop transformation or a framework that handles unstable or unbounded operators~\cite{chen2026soft,krebbekx2025lure,krebbekx2025nonlinear}.
\end{remark}

Next, we present the main result for discrete-time LTI systems.
\begin{theorem}[Discrete-time SRG stability theorem]\label{GFT for Discrete-time LTI systems}
    Consider the negative-feedback interconnection shown in Fig.~\ref{fig:feedback_system_diagram}.
    Let $H_1(z), H_2(z) \in \mathcal{RH}^{m\times m}_\infty$ be proper and stable discrete-time LTI systems.
    If, for all $\theta \in [0, \pi]$ and all $\tau \in (0,1]$,
    \begin{align}\label{discrete-time stability}
        \operatorname{SRG}(H_1(e^{j\theta}))^{-1}  \cap -\tau\operatorname{SRG}(H_2(e^{j\theta}))&= \varnothing,
    \end{align}
    then the homotopy interconnection obtained by replacing $H_2$ with $\tau H_2$ is well
    posed for every $\tau\in(0,1]$, and the closed-loop system is $\ell_2$-stable.
\end{theorem}
\begin{proof}
    Since $H_1$ and $H_2$ have real coefficients, 
    $H_i(e^{-j\theta})=\overline{H_i(e^{j\theta})}$,
    and $\operatorname{SRG}(\bar A)=\operatorname{SRG}(A)$ for any square matrix $A$.
    Hence,~\eqref{discrete-time stability} holds for every $\theta\in[0,2\pi)$.
    Fix arbitrary $\theta \in [0,2\pi)$ and $\tau \in (0,1]$, and define
    \begin{align*}
        A_1:=H_1(e^{j\theta}),\qquad A_2:=H_2(e^{j\theta}).
    \end{align*}
    The matrix $A_1$ may be singular.
    In that case, $0\in\operatorname{SRG}(A_1)$ and hence $\infty\in\operatorname{SRG}(A_1)^{-1}$,
    which never lies in the bounded set $-\tau\operatorname{SRG}(A_2)$.
    At the fixed frequency $\theta$, the homotopy feedback interconnection is described by
    \begin{align*}
        \hat{y} = A_1(\hat{u}-\tau A_2\hat{y}),
    \end{align*}
    or equivalently,
    \begin{align*}
        (I+\tau A_1 A_2)\hat{y} = A_1\hat{u} .
    \end{align*}
    We first show that condition~\eqref{discrete-time stability} implies
    \begin{align}
        \det(I+\tau A_1 A_2) \neq 0. \label{eq:det-nonzero}
    \end{align}
    Suppose, to the contrary, that $I+\tau A_1 A_2$ is singular.
    Then there exists $y\neq 0$ such that $(I+\tau A_1A_2)y=0$.
    Define $x:=-\tau A_2y$.
    Then
    \begin{align*}
        A_1x=y,
    \end{align*}
    so $x\neq 0$, since otherwise $y=A_1x=0$.
    Thus, $(x,y)$ is an input--output pair of $A_1$, and $(y,x)$ is an input--output pair
    of $-\tau A_2$, both with nonzero inputs.
    By the definition of the SRG and the scaling rule,
    \begin{align*}
        z_1&:=\frac{\|y\|}{\|x\|}e^{j\angle(x,y)}\in\operatorname{SRG}(A_1),\\
        z_2&:=\frac{\|x\|}{\|y\|}e^{j\angle(y,x)}\in\operatorname{SRG}(-\tau A_2)=-\tau\operatorname{SRG}(A_2).
    \end{align*}
    Since $\angle(y,x)=\angle(x,y)$, we have $z_2=1/\bar z_1\in\operatorname{SRG}(A_1)^{-1}$.
    Hence,
    \begin{align*}
        \operatorname{SRG}(A_1)^{-1}  \cap -\tau\operatorname{SRG}(A_2)&\neq \varnothing,
    \end{align*}
    which contradicts condition~\eqref{discrete-time stability}. 
    Hence,~\eqref{eq:det-nonzero} holds for every $\theta\in[0,2\pi)$ and every $\tau\in(0,1]$.

    Now define the homotopy family
    \begin{align*}
        F_\tau(z):=\det\big(I+\tau H_1(z)H_2(z)\big).
    \end{align*}
    At $\tau = 0$, $F_0(z) = 1$, so its winding number about the origin along the unit circle is zero.
    By~\eqref{eq:det-nonzero} and $F_0\equiv 1$,
    $F_\tau(e^{j\theta}) \neq 0$ for every point on the unit circle and every $\tau \in [0,1]$.
    Consequently, the family $F_\tau(e^{j\theta})$ defines a continuous homotopy in
    $\mathbb{C} \setminus \{0\}$,
    and its winding number about the origin is zero for every $\tau\in[0,1]$.

    Since $H_1$ and $H_2$ are proper and stable,
    $F_\tau$ is rational and analytic on $\{z:|z|\geq1\}\cup\{\infty\}$,
    with $F_\tau(\infty)=\det(I+\tau D_1D_2)$.
    For each $\tau\in[0,1]$, define $G_\tau(w):=F_\tau(1/w)$ for $w\neq0$ and
    $G_\tau(0):=F_\tau(\infty)$.
    Then $G_\tau$ is analytic in a neighborhood of the closed unit disk and has zero winding number about the origin along its boundary.
    By the argument principle, $G_\tau$ has no zeros in the closed unit disk.
    Thus, $F_\tau$ has no zeros in $\{z:|z|\geq1\}\cup\{\infty\}$ for every $\tau\in(0,1]$.
    In particular, $\det(I+\tau D_1D_2)\neq0$, i.e.,
    the homotopy interconnection is well posed for every $\tau\in(0,1]$.
    At $\tau=1$, $I+H_1H_2$ is analytic and nonsingular on $\{z:|z|\geq1\}\cup\{\infty\}$,
    so its inverse belongs to $\mathcal{RH}_\infty^{m\times m}$.
    Hence, the closed-loop transfer matrix $T_{cl}=(I+H_1H_2)^{-1}H_1$ from $u$ to $y$ belongs
    to $\mathcal{RH}_\infty^{m\times m}$.
    Therefore, by Parseval's identity,
    \begin{align*}
        \|y\|_{\ell_2} \leq \|T_{cl}\|_\infty\|u\|_{\ell_2},
    \end{align*}
    and the closed-loop system is $\ell_2$-stable.
\end{proof}

Exchanging scaling and geometric inversion as in Corollary~\ref{GFT inv for LTI systems}
gives an equivalent form of Theorem~\ref{GFT for Discrete-time LTI systems}.
\begin{corollary}\label{GFT inv for Discrete-time LTI systems}
    Under the assumptions of Theorem~\ref{GFT for Discrete-time LTI systems},
    if, for all $\theta \in [0, \pi]$ and all $\tau \in (0,1]$,
    \begin{align}\label{discrete-time stability inv}
        \tau\operatorname{SRG}(H_1(e^{j\theta})) \cap -\operatorname{SRG}(H_2(e^{j\theta}))^{-1} = \varnothing,
    \end{align}
    then the homotopy interconnection is well posed for every $\tau\in(0,1]$,
    and the closed-loop system is $\ell_2$-stable.
\end{corollary}
\begin{proof}
    Fix $\theta\in[0,\pi]$ and $\tau\in(0,1]$.
    Applying the argument in the proof of Corollary~\ref{GFT inv for LTI systems}
    with $\mathcal{S}_i:=\operatorname{SRG}(H_i(e^{j\theta}))$, $i=1,2$,
    shows that~\eqref{discrete-time stability inv} is equivalent to~\eqref{discrete-time stability}.
    The claim then follows from Theorem~\ref{GFT for Discrete-time LTI systems}.
\end{proof}

Theorem~\ref{GFT for Discrete-time LTI systems} establishes a frequency-wise stability
certificate by comparing the SRGs of the two subsystems at each frequency.
This formulation preserves frequency correspondence and therefore avoids intersections
arising from subsystem behaviors at different frequencies.

\section{Operator-Level SRG Stability Condition}\label{OperatorLevel}
We now derive an operator-level sufficient condition based on the SRGs of the causal
convolution operators induced by $H_1$ and $H_2$.
Because each operator SRG aggregates subsystem behavior across the entire frequency range,
the resulting condition is typically more conservative than the frequency-wise criterion
in Theorem~\ref{GFT for Discrete-time LTI systems}.
The closures of these operator SRGs are characterized in~\cite{shi2026computing}.

\begin{lemma}[Operator SRG closure~\cite{shi2026computing}]\label{lem:closure}
    Let $H\in\mathcal{RH}_\infty^{m\times m}$ be a discrete-time system,
    and let $T_d:\ell_2\to\ell_2$ be the causal convolution operator induced by $H$.
    Then
    \begin{align}
        f_{bk}(\mathrm{cl}\,\operatorname{SRG}(T_d))=\mathrm{co}\bigcup_{\theta\in[0,2\pi)}\mathcal{W}\big(f_{bk}(H(e^{j\theta}))\big). \label{eq:closure}
    \end{align}
    In particular, $\operatorname{SRG}(H(e^{j\theta}))\subseteq\mathrm{cl}\,\operatorname{SRG}(T_d)$ for every $\theta\in[0,2\pi)$.
\end{lemma}

\begin{proof}
    Identity~\eqref{eq:closure} is proved in~\cite[Thm.~1]{shi2026computing},
    and the inclusion follows from~\eqref{eq:closure} and Lemma~\ref{lem:matrix-srg}.
    For completeness, we give a direct proof of the inclusion.
    Fix $\theta_0\in[0,2\pi)$ and $x\in\mathbb{C}^m\setminus\{0\}$.
    For $N\geq1$, let $u_N(k):=N^{-1/2}xe^{jk\theta_0}$ for $0\leq k<N$, and $u_N(k):=0$ for $k\geq N$.
    Then $\|u_N\|=\|x\|$, and the boundary value of $u_N$ is $x\,d_N(\theta-\theta_0)$,
    where $d_N(\varphi):=N^{-1/2}\sum_{k=0}^{N-1}e^{-jk\varphi}$ and $|d_N|^2$ is the Fej\'er kernel.
    Since $H(e^{j\theta})$ is continuous and the Fej\'er kernel is an approximate identity,
    Parseval's identity gives $\|T_du_N\|\to\|H(e^{j\theta_0})x\|$ and
    $\langle u_N,T_du_N\rangle\to x^*H(e^{j\theta_0})x$ as $N\to\infty$.
    Hence, the SRG points generated by $u_N$ converge to those generated by $x$ under
    $H(e^{j\theta_0})$.
\end{proof}

\begin{corollary}\label{GFT for closure Discrete-time LTI systems}
    Under the assumptions of Theorem~\ref{GFT for Discrete-time LTI systems},
    let $T_{d,i}:\ell_2 \rightarrow \ell_2$, $i\in \{1,2\}$,
    be the causal convolution operator induced by $H_i$.
    If, for all $\tau\in(0,1]$,
    \begin{align}\label{discrete-time closure stability}
        \mathrm{cl}\,\operatorname{SRG}(T_{d,1})^{-1} \cap -\tau\,\mathrm{cl}\,\operatorname{SRG}(T_{d,2}) = \varnothing,
    \end{align}
    then the closed-loop system is $\ell_2$-stable.
    Equivalently, condition~\eqref{discrete-time closure stability} can be replaced by
    \begin{align}\label{discrete-time closure stability inv}
        \tau\,\mathrm{cl}\,\operatorname{SRG}(T_{d,1}) \cap -\mathrm{cl}\,\operatorname{SRG}(T_{d,2})^{-1} = \varnothing.
    \end{align}
\end{corollary}

\begin{proof}
    By Lemma~\ref{lem:closure},
    $\operatorname{SRG}(H_i(e^{j\theta}))\subseteq\mathrm{cl}\,\operatorname{SRG}(T_{d,i})$
    for every $\theta\in[0,2\pi)$ and $i\in\{1,2\}$.
    Since set inversion and real scaling preserve inclusions,
    \begin{align*}
        \operatorname{SRG}(H_1(e^{j\theta}))^{-1}&\subseteq\mathrm{cl}\,\operatorname{SRG}(T_{d,1})^{-1},\\
        -\tau\operatorname{SRG}(H_2(e^{j\theta}))&\subseteq-\tau\,\mathrm{cl}\,\operatorname{SRG}(T_{d,2}).
    \end{align*}
    Hence,~\eqref{discrete-time closure stability} implies~\eqref{discrete-time stability} for
    every $\theta\in[0,2\pi)$ and $\tau\in(0,1]$,
    and the claim follows from Theorem~\ref{GFT for Discrete-time LTI systems}.
    The exchanged form follows as in the proof of Corollary~\ref{GFT inv for LTI systems}.
\end{proof}

Corollary~\ref{GFT for closure Discrete-time LTI systems} compares two operator
SRGs that aggregate subsystem behavior across all frequencies and all $\ell_2$ inputs.
Consequently, an intersection in~\eqref{discrete-time closure stability} can arise
even when all same-frequency tests remain separated,
either from points generated at different frequencies in the two subsystems
or from the convex hull in~\eqref{eq:closure}.
This hull is taken in Beltrami--Klein coordinates and thus corresponds to a hyperbolic,
not Euclidean, convex hull in the SRG plane.
Hence, failure of the operator-level condition does not imply instability or failure of the
frequency-wise condition.
Since inversion and scaling preserve inclusions,
the corollary remains valid if the SRG closures are replaced by any sets containing them.

\section{Numerical Examples}\label{Examples}
\subsection{SISO Feedback Loops}
Consider the second-order plant and the first-order lag compensator
\begin{align*}
    H_1(z) = \frac{z}{(z-0.3)(z-0.5)}, \qquad H_2(z) = \frac{0.475z-0.4}{z-a}.
\end{align*}
The plant has poles at $0.3$ and $0.5$, 
and the compensator has a zero at $z\approx 0.842$ and a pole at $z=a$.
We consider $a=0.95$ (Example~1) and $a=0.98$ (Example~2).
Both subsystems are stable, and $H_1$ is strictly proper. 
Therefore, $D_1=0$, and the homotopy interconnection is well posed for every $\tau\in(0,1]$.
Closed-loop pole calculations confirm stability, 
with largest pole moduli of $0.8793$ and $0.8908$ for Examples~1 and~2, respectively.

We evaluate the frequency responses at $\theta=0$ and at $1000$ logarithmically spaced frequencies in $[10^{-3},\pi]$.
For a SISO subsystem, $\operatorname{SRG}(H_i(e^{j\theta}))$ consists of $H_i(e^{j\theta})$ and its complex conjugate.
The inverse controller SRG, $-\operatorname{SRG}(H_2(e^{j\theta}))^{-1}$, similarly consists of $-1/H_2(e^{j\theta})$ and its complex conjugate.
At each sampled frequency, we test~\eqref{discrete-time stability inv} by checking these inverse SRG points against the two radial segments traced by
$\tau\operatorname{SRG}(H_1(e^{j\theta}))$, $\tau\in(0,1]$.

\begin{figure}[t]
    \centering
    \includegraphics[width=\columnwidth]{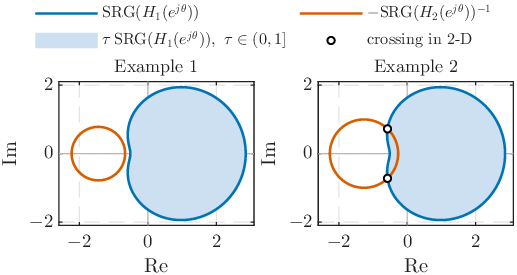}
    \caption{2-D SRGs of the SISO loop.}
    \label{fig:siso-2d}
\end{figure}

\begin{figure}[t]
    \centering
    \includegraphics[width=0.9\columnwidth]{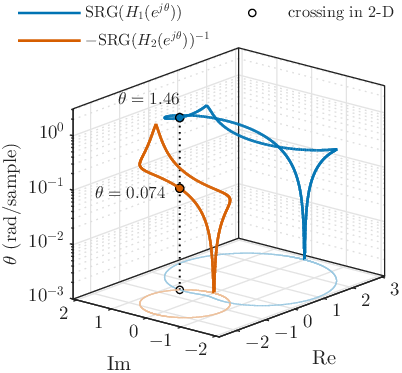}
    \caption{Frequency-wise SRGs of Example~2 ($a=0.98$).}
    \label{fig:siso-3d}
\end{figure}

Figs.~\ref{fig:siso-2d} and~\ref{fig:siso-3d} show the curves at $\tau=1$, corresponding to the original feedback interconnection.
For $\tau<1$, each plant SRG point is scaled toward the origin.
Therefore, $\tau=1$ gives the largest radial extent.
The shading in Fig.~\ref{fig:siso-2d} represents the union of the scaled plant sets over the displayed frequencies and $\tau\in(0,1]$.

In Example~1, Fig.~\ref{fig:siso-2d} shows that the inverse controller SRG lies entirely to the left of the shaded plant sets.
Consequently,~\eqref{discrete-time stability inv} is satisfied at every sampled frequency for all $\tau\in(0,1]$.

In Example~2, the SRG curves intersect at the circled points in Fig.~\ref{fig:siso-2d}.
Fig.~\ref{fig:siso-3d} resolves these intersections along the frequency axis.
The crossing pairs $H_1(e^{j\theta})$ at $\theta\approx1.46$ with $-1/H_2(e^{j\theta})$ at $\theta\approx0.074$.
By Lemma~\ref{lem:closure}, this intersection also belongs to the sets in~\eqref{discrete-time closure stability inv}, so the operator-level condition fails at $\tau=1$.

In contrast, at each sampled frequency, 
the inverse controller SRG remains disjoint from the scaled plant SRG for every $\tau\in(0,1]$.
Thus, the frequency-wise test succeeds on the grid for both examples, 
in agreement with the closed-loop pole calculations.
Example~2 shows how preserving frequency correspondence avoids a cross-frequency intersection that would otherwise make the operator-level test inconclusive.

\subsection{MIMO Feedback Loops}
Consider the $2\times2$ plant of~\cite{cordoba2025data} and the first-order controller
\begin{align*}
    H_1(z)&=\frac{1}{z-0.9048}
    \begin{bmatrix}0.09516&0.03807\\-0.02974&0.04758\end{bmatrix},\\
    H_2(z)&=\frac{c}{z-0.3}
    \begin{bmatrix}1&-0.3125\\0.4&0.5\end{bmatrix}.
\end{align*}
We consider $c=2$ (Example~3) and $c=10$ (Example~4).
Both subsystems are stable, and $H_1$ is strictly proper,
so the homotopy interconnection is well posed for every $\tau\in(0,1]$.
Closed-loop pole calculations give largest pole moduli of $0.7703$ and $1.1789$ for Examples~3 and~4, respectively.
Thus, the loop is stable in Example~3 and unstable in Example~4.
\begin{figure}[t]
    \centering
    \includegraphics[width=0.9\columnwidth]{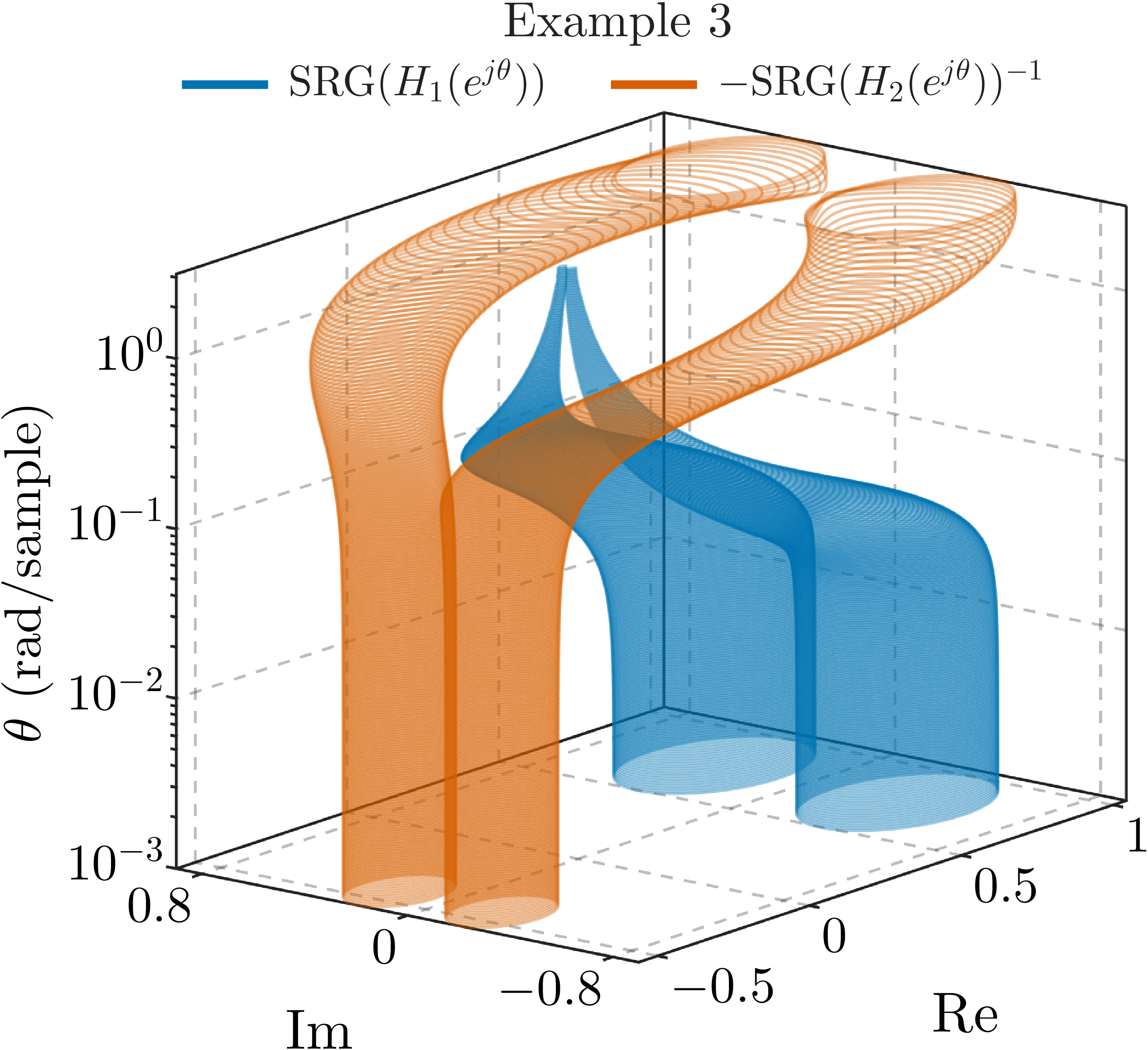}
    \caption{Frequency-wise SRGs of Example~3 ($c=2$).}
    \label{fig:mimo-3d-s}
\end{figure}

\begin{figure}[t]
    \centering
    \includegraphics[width=0.9\columnwidth]{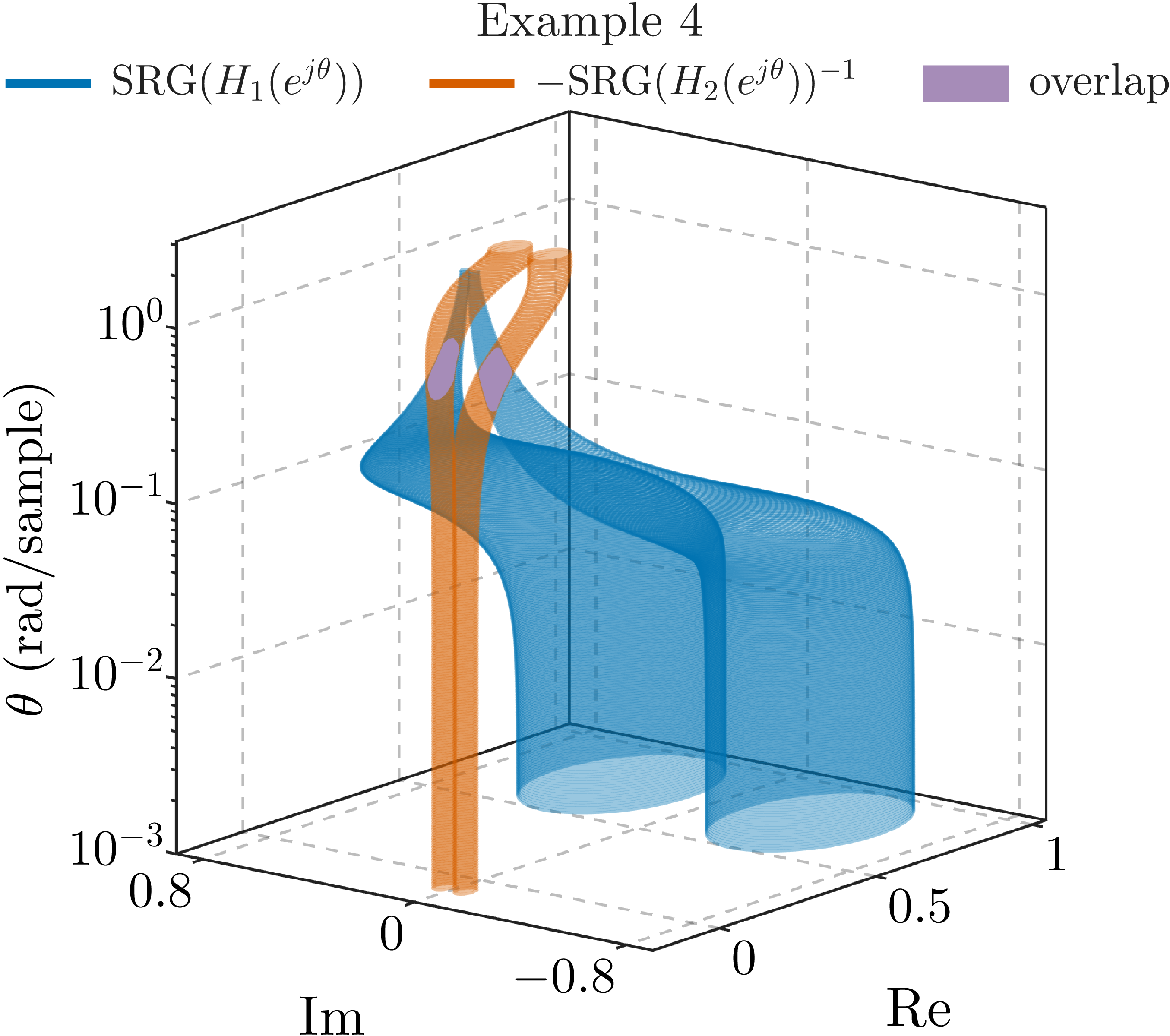}
    \caption{Frequency-wise SRGs of Example~4 ($c=10$).}
    \label{fig:mimo-3d-u}
\end{figure}

\begin{figure}[t]
    \centering
    \includegraphics[width=\columnwidth]{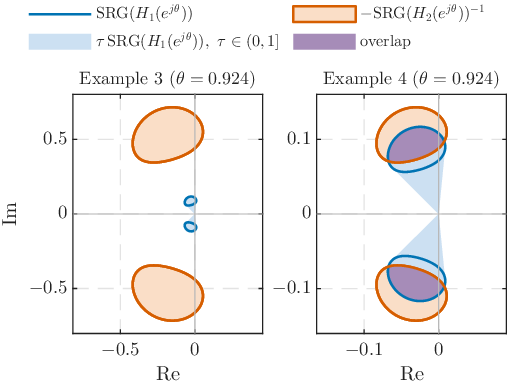}
    \caption{SRGs of the MIMO loops at $\theta=0.924$.}
    \label{fig:mimo-slices}
\end{figure}
We evaluate the frequency responses at $\theta=0$ and at $1000$ logarithmically spaced frequencies in $[10^{-3},\pi]$.
Unlike in the SISO case, each $\operatorname{SRG}(H_i(e^{j\theta}))$ is a region,
which we compute from Lemma~\ref{lem:matrix-srg}.
At each sampled frequency, we test~\eqref{discrete-time stability inv} by checking
$-\operatorname{SRG}(H_2(e^{j\theta}))^{-1}$ against the union of the scaled plant sets
$\tau\operatorname{SRG}(H_1(e^{j\theta}))$ over $\tau\in(0,1]$.

Fig.~\ref{fig:mimo-3d-s} shows the plant set $\operatorname{SRG}(H_1(e^{j\theta}))$
and the controller set $-\operatorname{SRG}(H_2(e^{j\theta}))^{-1}$ of Example~3
at $\tau=1$ along a logarithmic frequency axis.
At each grid frequency, the controller set is disjoint from the scaled plant sets for all $\tau\in(0,1]$,
so the condition of Corollary~\ref{GFT inv for Discrete-time LTI systems} holds on the full grid.
If frequency correspondence is ignored, however,
the controller set at high frequencies intersects the plant set at low frequencies.
Hence, the operator-level condition of Corollary~\ref{GFT for closure Discrete-time LTI systems} fails, 
although the loop is stable.

Fig.~\ref{fig:mimo-3d-u} shows the SRGs of Example~4, which is an unstable loop.
The two sets intersect over a range of frequencies, 
shown in purple in Fig.~\ref{fig:mimo-3d-u},
so the condition of Corollary~\ref{GFT inv for Discrete-time LTI systems} fails,
consistent with the instability of the loop.
To examine this failure at a single frequency,
we choose a representative frequency $\theta=0.924$.
Fig.~\ref{fig:mimo-slices} shows the SRGs of both examples at this frequency.
On the left (Example~3), the controller set is disjoint from the scaled plant sets.
On the right (Example~4), the controller set intersects the scaled plant sets,
so~\eqref{discrete-time stability inv} fails at this frequency.
In fact, such an intersection is guaranteed for an unstable loop:
by the contrapositive of Corollary~\ref{GFT inv for Discrete-time LTI systems},
\eqref{discrete-time stability inv} must fail for some $\theta$ and $\tau\in(0,1]$.

\section{Conclusions}\label{Conclusions}
This paper has presented a frequency-wise SRG condition for the feedback stability of proper,
stable discrete-time LTI systems.
We derived a corollary of the existing continuous-time theorem by moving geometric inversion from one subsystem SRG to the other. 
The same transformation yields two equivalent forms of the discrete-time stability condition, 
each applying geometric inversion to one subsystem SRG.
Both conditions guarantee well-posedness along the feedback-scaling homotopy and closed-loop $\ell_2$-stability.
These equivalent formulations operate directly on SRG sets and require no inverse-system realization.
The operator-level comparison shows that combining responses across frequencies and taking a hyperbolic convex hull can introduce additional conservatism.
Numerical examples with SISO and MIMO systems illustrate this effect: for stable loops, the SRGs remain disjoint at corresponding frequencies, despite occasional intersections across frequencies.
These results show that preserving frequency correspondence can reduce the conservatism of SRG-based stability analysis. 
In future work, we plan to extend the analysis to subsystems with integrators or unstable poles and to investigate SRG separation as a stability margin for controller design.

\bibliographystyle{IEEEtran}
\bibliography{main_ECC}

\end{document}